\documentclass{article}

\PassOptionsToPackage{numbers}{natbib}

\usepackage[final]{neurips_2024}

\usepackage[utf8]{inputenc} 
\usepackage[T1]{fontenc}    
\usepackage{hyperref}       
\usepackage{url}            
\usepackage{booktabs}       
\usepackage{amsfonts}       
\usepackage{nicefrac}       
\usepackage{blindtext}
\usepackage{microtype}      
\usepackage{xcolor}         
\usepackage{amsmath}
\usepackage{physics}
\usepackage{graphicx}
\usepackage{subcaption}
\usepackage{float}
\usepackage{amssymb}
\usepackage{mathrsfs}
\usepackage{relsize}
\usepackage{bigints}
\usepackage{algorithm,algcompatible}

\newcommand{\R}{\mathbb{R}}

\newcommand{\E}{\mathbb{E}}

\algnewcommand\INPUT{\item[\textbf{Input:}]}%
\algnewcommand\OUTPUT{\item[\textbf{Output:}]}%

\title{Accelerated Image-Aware Generative Diffusion Modeling}

\author{%
  Tanmay~Asthana\\
  Department of Electrical and Computer Engineering\\
  North Carolina State University\\
  Raleigh, NC 27606 \\
  \texttt{tasthan@ncsu.edu} \\
  \And
  Yufang~Bao\\
  Department of Mathematics and Computer Science\\
  Fayetteville State University\\
  Fayetteville, NC 28301
  \And
  Hamid~Krim\\
  Department of Electrical and Computer Engineering\\
  North Carolina State University\\
  Raleigh, NC 27606 \\
 \\
}

\begin{document}

\maketitle

\begin{abstract}
  We propose in this paper an analytically new construct of a diffusion model whose drift and diffusion parameters yield an exponentially time-decaying Signal to Noise Ratio in the forward process. In reverse, the construct cleverly carries out the learning of the diffusion coefficients on the structure of clean images using an autoencoder. The proposed methodology significantly accelerates the diffusion process, reducing the required diffusion time steps from around 1000 seen in conventional models to 200-500 without compromising image quality in the reverse-time diffusion. In a departure from conventional models which typically use time-consuming multiple runs, we introduce a parallel data-driven model to generate a reverse-time diffusion trajectory in a single run of the model. The resulting collective block-sequential generative model eliminates the need for MCMC-based sub-sampling correction for safeguarding and improving image quality, to further improve the acceleration of image generation. Collectively, these advancements yield a generative model that is an order of magnitude faster than conventional approaches, while maintaining high fidelity and diversity in generated images, hence promising widespread applicability in rapid image synthesis tasks.
\end{abstract}

\section{Introduction}
Generative diffusion models (GMs) have recently emerged as powerful tools for image modeling and numerous other applications \cite{DPM, NCSNs, DDPM, Scored}, offering exceptional fidelity and generative diversity \cite{Survey}. In contrast to existing generative models, like generative adversarial networks (GANs) \cite{GANS} and variational autoencoders (VAEs) \cite{VAES}, GMs are more stable in training and less sensitive to hyper-parameter selection\cite{VAES1}.

Generative diffusion approaches gradually corrupt image data with increasingly random noise in the forward diffusion steps. The noise removal\footnote{This may equivalently be viewed as learning the statistical structure of the data for given additive white noise.} is progressively learned in the reverse diffusion to recover the desired image data to best match its initial distribution. 
Initially used in diffusion probabilistic models \cite{DPM}, the score-based
generative models (SGMs) - represented by Noise-Conditional Score Networks (NCSNs) \cite{NCSNs}  and the denoising diffusion probabilistic models (DDPMs) \cite{DDPM} - were successfully used to achieve high quality image samples through denoising score matching and annealed Langevin dynamics. The aforementioned generative diffusion models are fundamentally based on continuous stochastic differential equations (SDEs) \cite{Scored}, with a continuous forward diffusion model is defined to add random noise to an image according to some selected drift and diffusion parameters, and a reverse diffusion \cite{Anderson} is run by a reverse-time SDE governed by a score-based model to be learned by deep neural networks. 

 While effective, the performance of conventional diffusion models entails a slow convergence, with a quality image generation requiring a large number of time steps in turn, leading to an increased computational complexity. To this end, much effort \cite{FAST, FAST1, FAST2, FAST3, FAST4, FAST5, FAST6} has been dedicated to reducing this lengthy process and to improving the quality for prediction-correction methods by Markov chain Monte Carlo (MCMC) subsampling and modified Langevin dynamics.  
  In this paper, we propose an alternative approach and use insights from statistical mechanics of particles to account for local (pixel level) SNR in driving the microscopic dynamics of the GM diffusion. 
  In so doing, our novel diffusion model leverages the structure of clean image data to learn the drift and diffusion parameters at a microscopic level. Specifically, these parameters increase the rate of degradation according to image pixel- SNR in contrast to the uniform regime of existing SDE-based models. This is inspired by the well known water pouring algorithm paradigm \cite{WATER} employed in multi-channel communication systems. The water pouring algorithm allocates power to a channel in accordance with the noise-level experienced in that channel, with more degraded channels getting more power. Intuitively, one may interpret the  macroscopic forward diffusion as a parallel (bundle) process of  microscopic forward diffusion processes occurring on individual pixels in parallel. The forward diffusion scheduling is dependent, as detailed later, on the initial clean pixel values. We demonstrate that by employing this pixel based scheduling  strategy, we can achieve the target goal of reaching isotropic Gaussian distribution on all the pixels much faster than the conventional pixel agnostic diffusion scheduling. 

With such an image-aware forward diffusion in hand, we proceed with an autoencoder to learn the combined diffusion schedule across all the pixels of a noisy image. This learned schedule is subsequently used in a data-driven reverse-time diffusion model to generate the complete trajectory of a reverse-time diffusion. While conventional models generate the reverse trajectory one step at a time, we leverage the structural information learned in the scheduling strategy to generate the whole reverse-time diffusion path in one go without compromising on the over-all generated image quality. As a result of this strategy, we are able to accelerate the reverse-time diffusion process by nearly an order of magnitude. 

\section{Background}

subsection{Forward diffusion}
Associating to a data sample $\vb{x}_0 \in \R^d$ distributed as $\vb{x}_0 \sim q(\vb{x}_0)$, a common forward diffusion \cite{DPM} is tantamount to defining, a Markov chain of samples $\vb{x}_1,...,\vb{x}_T$ such that:
\begin{equation}
   q(\vb{x}_1,...,\vb{x}_T|\vb{x}_0) = \prod_{i=1}^Tq(\vb{x}_i|\vb{x}_{i-1}),\,\,\, q(\vb{x}_i|\vb{x}_{i-1}) = \mathcal{N}(\sqrt{1-\beta_i}\vb{x}_{i-1},\beta_i\textbf{I}),\,\,\, i \in \{1,...,T\}, 
    \label{equ:Gaussian}
\end{equation}
where $\beta_i\in (0,1),\,\,\forall \,\, i \in \{1,...,T\}$, is an increasing scalar schedule $i$ from a very small positive value towards $1$ in $T$ steps. For a sufficiently large, $T$, a well behaved increasing schedule $\beta_i$ ensures that $\vb{x}_T \sim \mathcal{N}(\textbf{0},\textbf{I})$ .
In relation to Eqn.~\ref{equ:Gaussian}, a  forward diffusion of this form may follow as a discretized process,
\begin{equation}
    \vb{x}_{i+1} = \sqrt{\alpha_i} \vb{x}_i + \sqrt{1-\alpha_i} \vb{\epsilon}_i,\,\,\, \vb{\epsilon}_i \sim \mathcal{N}(\textbf{0},\textbf{I}),
    \label{equ:iterative_fwd}
\end{equation}

where $\alpha_i = 1 - \beta_i$, the first addend is a drift term of $\vb{x}_{i+1}$, and the second is its diffusion term. An iterative form of Eqn.~\ref{equ:iterative_fwd} in terms of the initial sample value $\vb{x}_0$, can also be obtained  as a reparameterized equation\cite{VAES},
\begin{equation}
    \vb{x}_{i+1} = \sqrt{\overline{\alpha}_i} \vb{x}_0 + \sqrt{1-\overline{\alpha}_i} \Tilde{\vb{\epsilon}}_i,\,\,\, \Tilde{\vb{\epsilon}}_i \sim \mathcal{N}(\textbf{0},\textbf{I}),
\end{equation}
where $\overline{\alpha}_i = \prod_{k=1}^i\alpha_k$ and $\Tilde{\vb{\epsilon}}_i$ is a linear combination of $\vb{\epsilon}_k \sim \mathcal{N}(\textbf{0},\textbf{I}),\,\,\,k \in \{1,...,i\}$, such that $\Tilde{\vb{\epsilon}}_i \sim \mathcal{N}(\textbf{0},\textbf{I})$.
Note that  $\beta_i$ controls the diffusion rate of the process, and a number of its variations have been used including a linear profile \cite{DDPM}, a harmonic \cite{DPM} and a squared cosine dependency \cite{DDIM}. In all these cases, a simple scalar $\beta_i$ is invariably used for each element of the vector $\vb{x}_i = [x^1_i,\cdots,x^j_i,\cdots, x^d_i] \in \R^d$, where $d$ is the number of pixels in an image. 

\subsection{Sampling for reverse-time diffusion}
While Eqn.~\ref{equ:iterative_fwd} implies $q(\vb{x}_i|\vb{x}_{i-1})$ is an explicitly known conditional Gaussian distribution for  $\vb{x}_i$, the posterior distribution $q(\vb{x}_{i-1}|\vb{x}_i)$ is unknown.  The probability of $\vb{x}_{i-1}$ conditioned on $\vb{x}_i$ and $\vb{x}_0$
can, however, be explicitly expressed using Bayes rule \cite{DDPM} as

\begin{equation}
    \begin{split}
        q(\vb{x}_{i-1}|\vb{x}_i,\vb{x}_0) &= \mathcal{N}(\hat{\mu}(\vb{x}_i,\vb{x}_0),\Tilde{\beta}_i\textbf{I}) = \mathcal{N}(\Tilde{\mu}(\vb{x}_i,\Tilde{\vb{\epsilon}}_i),\Tilde{\beta}_i\textbf{I}),\\
        \hat{\mu}(\vb{x}_i,\vb{x}_0) &= \dfrac{\sqrt{\overline{\alpha}_{i-1}}\beta_i}{1-\overline{\alpha}_i}\vb{x}_0 + \dfrac{\sqrt{\alpha_i}(1-\overline{\alpha}_{i-1})}{1-\overline{\alpha}_i}\vb{x}_i,\\
        \Tilde{\mu}(\vb{x}_i,\Tilde{\vb{\epsilon}}_i) &=  \dfrac{1}{\sqrt{\alpha_t}}\Bigg(\vb{x}_i - \dfrac{1-\alpha_i}{\sqrt{1-\overline{\alpha}_i}}\Tilde{\vb{\epsilon}}_i\Bigg),\\
        &\Tilde{\beta}_i = \dfrac{1-\overline{\alpha}_{i-1}}{1-\overline{\alpha}_i}\beta_i.
    \end{split}
    \label{equ:posterior}
\end{equation}

At inference time (i.e. during reverse-time diffusion) $\vb{x}_0$ is not known. The equivalence between the means  
$\hat{\mu}(\vb{x}_i,\vb{x}_0)$ and $\Tilde{\mu}(\vb{x}_i,\Tilde{\vb{\epsilon}}_i)$, helped all models based on the Denoising Diffusion Probabilistic Model (DDPM) \cite{DDPM},\cite{DPM},\cite{DDIM},\cite{DiffTrans} approximate $q(\vb{x}_{i-1}|\vb{x}_i,\vb{x}_0)$ using $p_{\theta}(\vb{x}_{i-1}|\vb{x}_{i}) = \mathcal{N}\Big(\Tilde{\mu}\Big(\vb{x}_i,\vb{\varepsilon}_{\theta}(\vb{x}_i,i)\Big),\Tilde{\beta}_i\Big)$, where $\vb{\varepsilon}_{\theta}(\vb{x}_i,i)$ is a neural network model for noise removal added in the forward diffusion step. To learn to predict the mean $\Tilde{\mu}(.)$, a Variational Lower Bound (VLB) on the negative log likelihood $p_{\theta}(\vb{x}_0)$ is maximized. This VLB is given as:

  \begin{equation}
    \begin{split}
          &L_{\text{VLB}} = \E_q\Big[D_{\text{KL}}\Big(q(\vb{x}_T|\vb{x}_0)||p_{\theta}(\vb{x}_T)\Big)\\
         &+ \sum_{i=2}^TD_{\text{KL}}\Big(q(\vb{x}_{i-1}|\vb{x}_i,\vb{x}_0)||p_{\theta}(\vb{x}_{i-1}|\vb{x}_{i})\Big) - \log{\Big(p_{\theta}(\vb{x}_{0}|\vb{x}_{1}\Big))}\Big].
    \end{split}
\label{equ:VLB}
 \end{equation}

$\overline{\alpha}_i$, $\beta_i$ and the neural network output $\vb{\varepsilon}_{\theta}(\vb{x}_i,i)$), enables reverse-time diffusion trajectory construction using the following, 
\begin{equation}
    \vb{x}_{i-1} = \Tilde{\mu}(\vb{x}_i,\vb{\varepsilon}_{\theta}(\vb{x}_i,i)) + \sqrt{\Tilde{\beta}_i} \sigma_i,\,\,\, \sigma_i \sim \mathcal{N}(\textbf{0},\textbf{I}),\,\,\,i \in \{T,\cdots,1\}.
    \label{equ:reverse_samples_eps}
\end{equation}
This perspective has led to various generative diffusion models to sample from the posterior distribution.

Additionally, a continuous version of Eqn. \ref{equ:iterative_fwd} was given in  \cite{Scored} for $T \rightarrow \infty$ yielding a forward-time Stochastic Differential Equation (SDE).
\begin{equation}
    d\vb{x}_{t} = \vb{f}(\vb{x}_t,t)dt + g(\vb{x_t},t)d\textbf{w},\,\,\, t \in (0,1).
        \label{equ:continous}
\end{equation}
Relying on \cite{Anderson} and \cite{HYAV},  the corresponding reverse-time SDE can be built via learning the so-called score function $\nabla_{\vb{x}_t}\log{(q(x_t))}$  iterately through through a neural network.

Two main neural network architectures are used for learning tasks: U-Net \cite{NCSNs},\cite{DDPM},\cite{Scored},\cite{DDIM} and Transformer-based models \cite{DiffTrans},\cite{Stable_Diff}. The U-Net \cite{UNET}, popular for semantic segmentation tasks, uses a downsampling encoder and an upsampling decoder, with feature maps from the encoder concatenated with inputs of the decoder at different resolutions. Upsampling is a sparse operation. A good prior from earlier stages aids the decoder to better represent the localized features. Newer models use Transformer architectures, which work on lower-dimensional latent encodings instead of images, offering higher generation quality despite being more computationally intensive \cite{DiffTrans}.

\section{Methodology: Image Aware Diffusion}

\subsection{Motivation}
Our forward diffusion objective is a standard isotropic multi-dimensional normal distribution for the data, iteratively attained by attenuating/enhancing the drift/diffusion terms. The evolution of the various data pixels to $N({\mathbf 0},{\mathbf I})$ is clearly dependent on their individual initial values, and equalizing the group-evolution is reasonable.  The water pouring algorithm employed in multi-channel communication systems\cite{WATER}, similarly addresses assignment of signal power distribution across frequency channels with different ambient noise powers to maximize SNR. Faced with our objective of all pixels simultaneously achieving a standard normal distribution over a certain time interval (the total number of steps, $T$), it makes sense to diffuse higher-valued pixels at a much faster rate than those of lower-valued pixels. This clearly requires pixel-value-dependent drift and diffusion parameters, allowing e.g., a relatively higher power noise for higher-valued pixels and vice-versa. 

This also implies that our group-diffusion process, will require a vector  $\boldsymbol{\beta}_i \in \R^d$, whose elements are different for different pixels $x^j_i$, in $\vb{x}_i = [x^1_i,\cdots,x^j_i,\cdots, x^d_i] \in \R^d$. We propose here a carefully chosen image-aware scheduling forward process.  Consequently the reverse process  can converge much faster.

\subsection{Redefining forward diffusion}
\subsubsection{Definitions}
\label{sec:def}
We define image scale $\vb{x}_{\delta} \in \R^d$ as
\begin{equation}
    \vb{x}_{\delta} = \mathlarger{e}^{\mathlarger{-\gamma\vb{x}_0}},
    \label{eqn:scale}
\end{equation}
where $\gamma$ is a scalar hyperparameter such that $\gamma < T,\,\,\,\gamma >> x^j_0,\,\,\forall\,\,j\in\{1,\cdots,d\}$ and exponentiation is done elementvise.

The diffusion schedule parameters now become vectors, redefined as
\begin{equation}
    \boldsymbol{\alpha}_i={\bf 1}-\boldsymbol{\beta}_i = \vb{x}_{\delta}^{1/T}.
\end{equation}
This allows the diffusion schedule to vary for all the pixels $x^j_i,\,\, j\,\,\in \{1,...,d\}$. The resulting reparametrized forward step dependent on $\vb{x}_0$ can be written as
\begin{equation}
    \vb{x}_i = \sqrt{\overline{\boldsymbol{\alpha}}_i}\odot \vb{x}_{0} + \sqrt{1-\overline{\boldsymbol{\alpha}}_i}\odot \Tilde{\vb{\epsilon}}_i,\,\,\,\overline{\boldsymbol{\alpha}}_i =  \mathlarger{e}^{\mathlarger{-\gamma\,\,i\,\vb{x}_0}/T},\,\,\, i \in \{0,1,...,T\}
    \label{equ:exponential_reparam}
\end{equation}
where $\odot$ is an element-wise multiplication.

We further assume that all $x^j_0$ are normalized to the range $(0,1]$, with small scalar value added to all $x^j_0$, so that none of the resulting pixels is exactly 0 to ensure that $\mathlarger{e}^{\mathlarger{-\gamma\,\,i\,\, \vb{x}_0/T}}$ vary with $i$. From Eqn. \ref{equ:exponential_reparam}, it is clear that as $i$ increases from 0 to $T$, the drift term decreases exponentially from $\vb{x}_0$ to a very small value, while the noise variance increases exponentially from 0 to 1.

Restricting $\gamma < T$ ensures that the $\mathlarger{e}^{\mathlarger{-\gamma\,\,i\,\, \vb{x}_0/T}}$ terms are not vanishingly small for all $i$, hence preventing an exceedingly fast decay of image data adversely impacting the Markovian property of the sample chain. Simultaneously, $\gamma >> x^j_0$ ensures that the pixel density drift to a very small value as $i$ approaches $T$. These two conditions appear to be contradictory to each other and a compromise between them has to be established for each experiment. In our experiments for CIFAR10 dataset images of $32 \times 32$ resolution, we fixed $T =200$  and $\gamma =20$. For CelebA dataset images of $128 \times 128$ resolution, we fixed these values to 500 and 50 respectively. 

To analyze the trajectory of our diffusion model in time, we substitute discrete ratio $i/T$ with a continuous variable $t \in [0,1]$, by letting $T \rightarrow \infty$. As a result the discrete time-step Eqn. \ref{equ:exponential_reparam} becomes a continuous time one:

\begin{equation}
    \vb{x}_t = \sqrt{\overline{\boldsymbol{\alpha}}_t}\odot \vb{x}_{0} + \sqrt{1-\overline{\boldsymbol{\alpha}}_t}\odot \Tilde{\vb{\epsilon}}_t,\,\,\,\overline{\boldsymbol{\alpha}}_t =  \mathlarger{e}^{\mathlarger{-\gamma\,\,t\,\vb{x}_0}/T},\,\,\, t \in [0,1].
    \label{equ:exponential_reparam_cont}
\end{equation}

Using Eqn. \ref{equ:exponential_reparam_cont}, we can calculate the time differentiable SNR at pixel $x^j_0$ as:
\begin{equation}
    \text{SNR}(j,t) = \dfrac{(x^j_0)^2}{e^{ \gamma\,\,t\,\, x^j_0}-1}
    \label{equ:cont_SNR}
\end{equation}

Form Eqn.  \ref{equ:cont_SNR}, it is clear that the SNR of any pixel decreases exponentially with time.

\subsubsection{Proposition 1}
For our diffusion starting from clean pixels, $x^j_0 < x^k_0$, there exists a $t_{\delta}$,
\begin{equation}
    \frac{d\,\text{SNR}(k,t)}{dt} < \frac{d\,\text{SNR}(j,t)}{dt} < 0,\,\, \forall t\,\,\in [0,t_{\delta}]
\end{equation}

The proof of this proposition can be found in Appendix \ref{Appendix:snr}. Thus, our choice of $\boldsymbol{\alpha}_i$ makes pixels with high values experience faster SNR reduction.

\subsubsection{Proposition 2}
\label{Proposition2}
For a linearly varying conventional DDPM schedule of the form, $\beta_t = at,\,\,\,t \in [0,1]$, the expected trajectory is $\mathlarger{\chi}_{\text{C}}(t) = \E[\vb{x}_t] = \vb{x}_0e^{-\frac{at}{2}^2}$. For the diffusion model defined as per Eqn - \ref{equ:exponential_reparam_cont}, the expected trajectory is $\mathlarger{\chi}_{\text{N}}(t)=\E[\vb{x}_t] = \vb{x}_0\odot e^{- \frac{\gamma\vb{x}_0 t}{2}}$. Choosing 
$\gamma x^j_0 > at ,\,\,\,\forall\,\,j\in \{1,\cdots,d\},\,\,\, t \in (0,1]$ ensures
\begin{equation}
    \mathlarger{|}\dfrac{\mathlarger{\chi}_{\text{C}}(t)}{dt}\mathlarger{|} < \mathlarger{|}\dfrac{\mathlarger{\chi}_{\text{N}}(t)}{dt}\mathlarger{|}
\end{equation}

The proof of this proposition can be found in Appendix \ref{Appendix:rates}. Thus, by carefully choosing $\gamma$, we can get faster convergence from our model in comparison to conventional DDPM model.

\begin{figure}[ht]
\centering
\begin{subfigure}[t]{0.48\textwidth}
\includegraphics[width=1\linewidth, height=7cm]{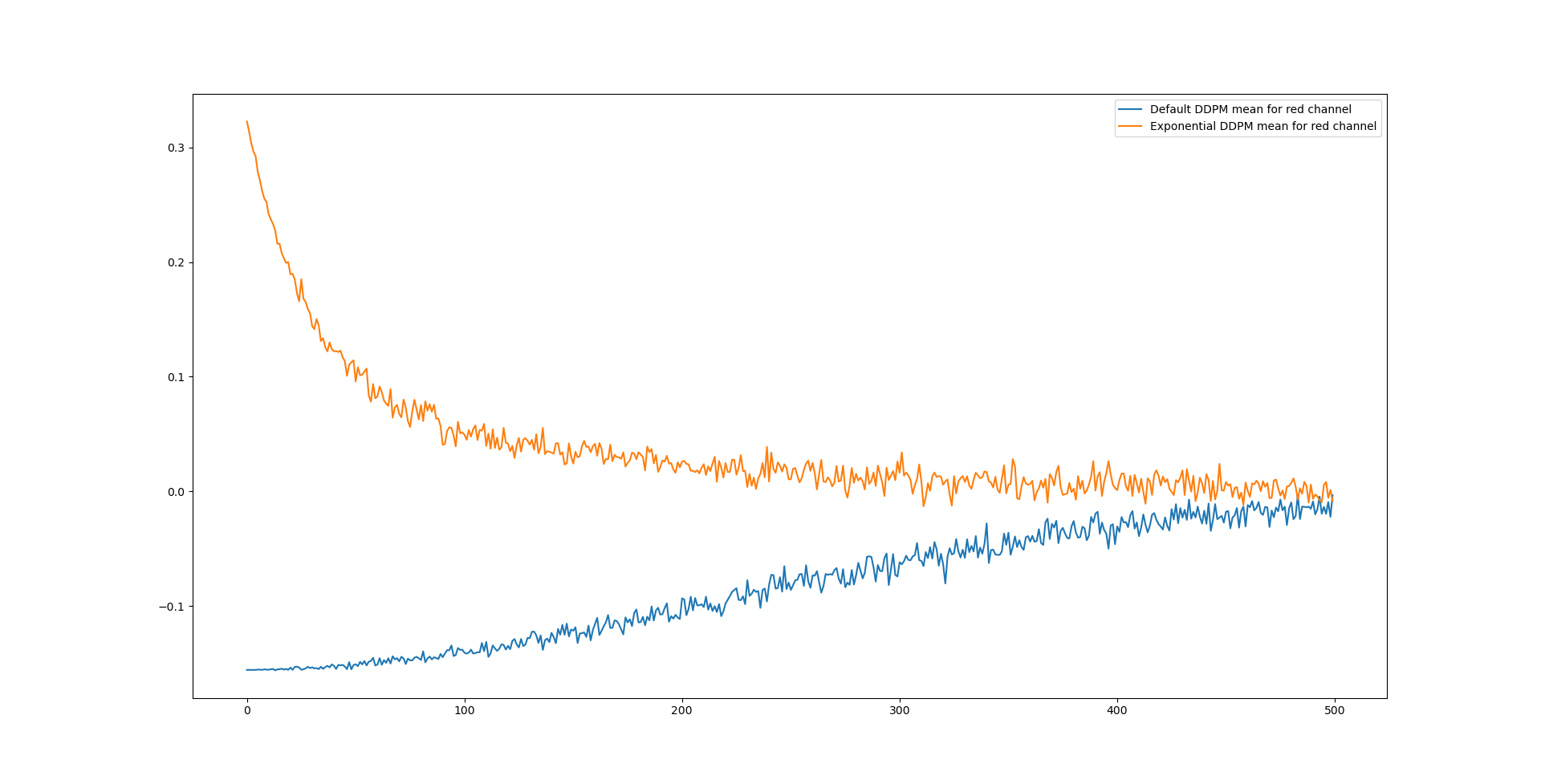}
\caption{Pixel mean progression}
\end{subfigure}
\begin{subfigure}[t]{0.48\textwidth}
\centering
\includegraphics[width=1\linewidth, height=7cm]{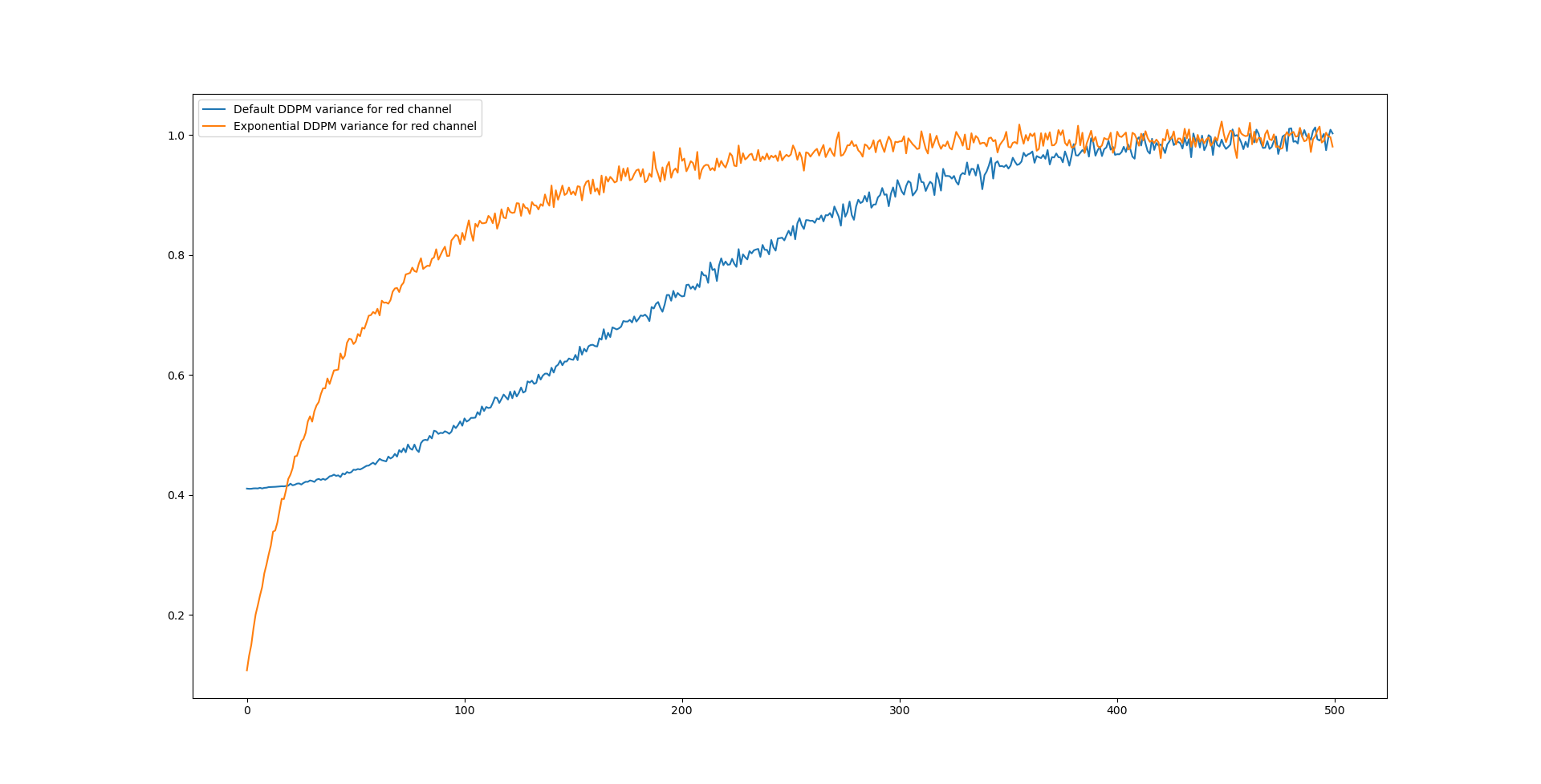}
\caption{Pixel variance progression}
\end{subfigure}
\caption{Comparison of pixels mean (left) and variance (right) progression in forward diffusion trajectory of a single color channel (red in this case) over time of the conventional DDPM (blue) vs. our model (orange). 
}
\label{fig:mean_var_comp}
\end{figure}

Fig.  \ref{fig:mean_var_comp} evidently demonstrates that the mean and variance of our new method converge to target values (0 and 1 respectively) much faster than the conventional diffusion model. 
Here the total number steps is kept constant to 500 steps. The $\beta_i$ for conventional DDPM  vary linearly from $10^{-4}$ to 0.02. The example image is a sample from the CelebA dataset that has a resolution of $128 \times 128$. The means and variances calculated in this experiment are empirical ones. We have assumed that as the number of pixels in the image is large enough (specifically $128 \times 128 = 16,384$), the empirical statistics would match theoretical statistics with high probability. Note that

The success of our algorithm requires to estimate the image scale $\vb{x}_{\delta} = e^{-\gamma\vb{x}_0}$,
As a result of exponentiation and large value of $\gamma$, image $\vb{x}_{\delta}$ lacks fine details. Consequently, it is easier to get its estimate from a noisy image $\vb{x}_i$ using any denoising based method. We employ a denoising autoencoder,$s_{\theta}(\vb{x}_i,i)$ for this purpose, such that $s_{\theta}(\vb{x}_i,i) = \hat{\vb{x}}_{\delta} \approx  \vb{x}_{\delta}$. For generative tasks, we use a variational autoencoder instead of a denoising one.

\begin{figure}[ht]
\centering
\begin{subfigure}{0.48\textwidth}
\includegraphics[width=1\linewidth, height=5cm]{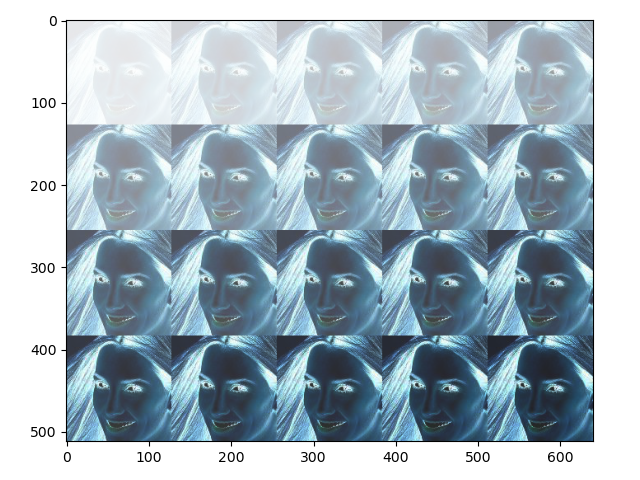}
\caption{Real $\overline{\alpha}_i$}
\end{subfigure}
\begin{subfigure}{0.48\textwidth}
\centering
\includegraphics[width=1\linewidth, height=5cm]{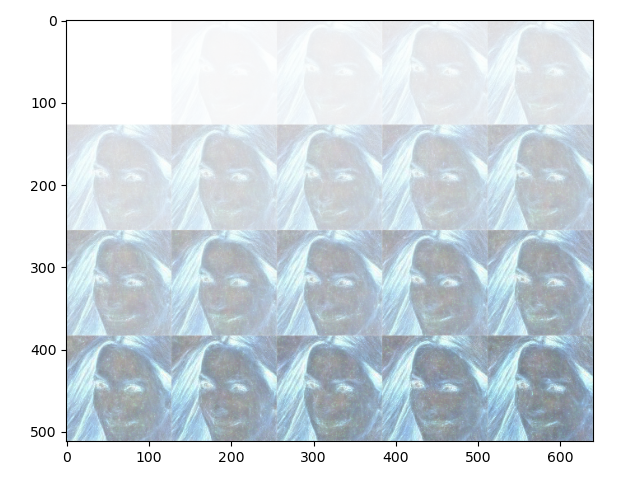}
\caption{Estimated $\overline{\alpha}_i$}
\end{subfigure}
\caption{Real $\overline{\alpha}_i$ and estimated $\overline{\alpha}_i$ (calculated from the image scale estimate, $\hat{\vb{x}}_{delta}$ for i ranging from 1 to 20.
}
\label{fig:alphabar_comp}
\end{figure}

Once $\vb{x}_{\delta}$  estimated, the approximations of the factors 
$\boldsymbol{\alpha_i}^j$ and $\overline{\boldsymbol{\alpha}}_i^j$  can also be readily calculated as they are simply elements from scaled exponents of $\vb{x}_{\delta}$ and element-wise used in Eqn. \ref{equ:posterior}. Fig. \ref{fig:alphabar_comp} shows the comparison of the real vs estimated $\overline{\boldsymbol{\alpha}}_i$ values. Fig. \ref{fig:scale_comp} in Appendix \ref{Appendix:scale_fig} show the comparison of the real vs estimated scales as well. The structural similarity (SSIM) index between real and estimated images was found to be in the range of 0.86 to 0.99 for both scale and $\overline{\boldsymbol{\alpha}}_i$ (the higher the better, with 1 signifying perfect similarity) across different values of $i$.

Choosing $\vb{x}_{\delta}$ as the autoencoder output instead of $\overline{\boldsymbol{\alpha}}_i$ simplifies the training
as it reduces the need to re-train the model for different $i$, which simplifies the architecture of the autoencoder, allowing us to model the autoencoder using a U-Net with lesser complexity. Details of its architecture are in Appendix \ref{appendix:arch}. From $\vb{x}_{\delta}$, we can also try to approximate the original image by taking the inverse of Eqn. \ref{eqn:scale}. However, this results in poor approximation, as shown in Fig.  \ref{fig:img_from_img_delta}. The structural similarity in this case was merely in the range of 0.23 to 0.42 across various values of $i$. Even tiny approximation errors in $\hat{\vb{x}_{\delta}}$ result in larger errors in the estimation of the clean image. Fig.  \ref{fig:rev_traj_real} shows as a toy example the last 20 steps of reverse trajectory using $\hat{\vb{x}}_{\delta}$, $\Tilde{\epsilon}_i$ and Eqn. \ref{equ:reverse_samples_eps}. 

\subsection{Reverse-time diffusion modelling}
We are ready for our reverse-time diffusion once $\vb{x}_{\delta}$, $\boldsymbol{\beta}_i$, $\boldsymbol{\alpha}_i$ and $\overline{\boldsymbol{\alpha}}_i$  can be estimated as shown in the previous subsection. However, the small errors in the scale and scheduling estimates of the autoencoder result in larger errors in the generated reverse trajectory. To compensate for these estimation errors, we introduce a modified U-net based neural network to generate the complete trajectory of the reverse-time diffusion process.

In conventional designs the same trained network is used to generate the reverse trajectory samples by using previously generated sample and next time-step positional encoding as input. This gives us evidence that the architecture has enough capacity to process the semantic information hidden in the noisy image at any time-step. We introduce following modifications to the conventional U-net based models:
\begin{enumerate}
    \item We also fuse (by addition to feature maps) information of the learned image scale $\hat{\vb{x}}_{\delta}$ predicted from the image scale autoencoder $s_{\theta}(x_i,i)$.
    \item We modify the structure of the last layer of the decoder to predict the additive noise for all the preceding time-steps ($\{i-1,\cdots,1\}$) in different channels of the last layer. This way our model has higher complexity than preceding models, but only a single execution of the trained model is required to get the clean image $\vb{x}_0$.
\end{enumerate}

\begin{figure}[ht]
\centering
\includegraphics[width=1\linewidth, height=6cm]{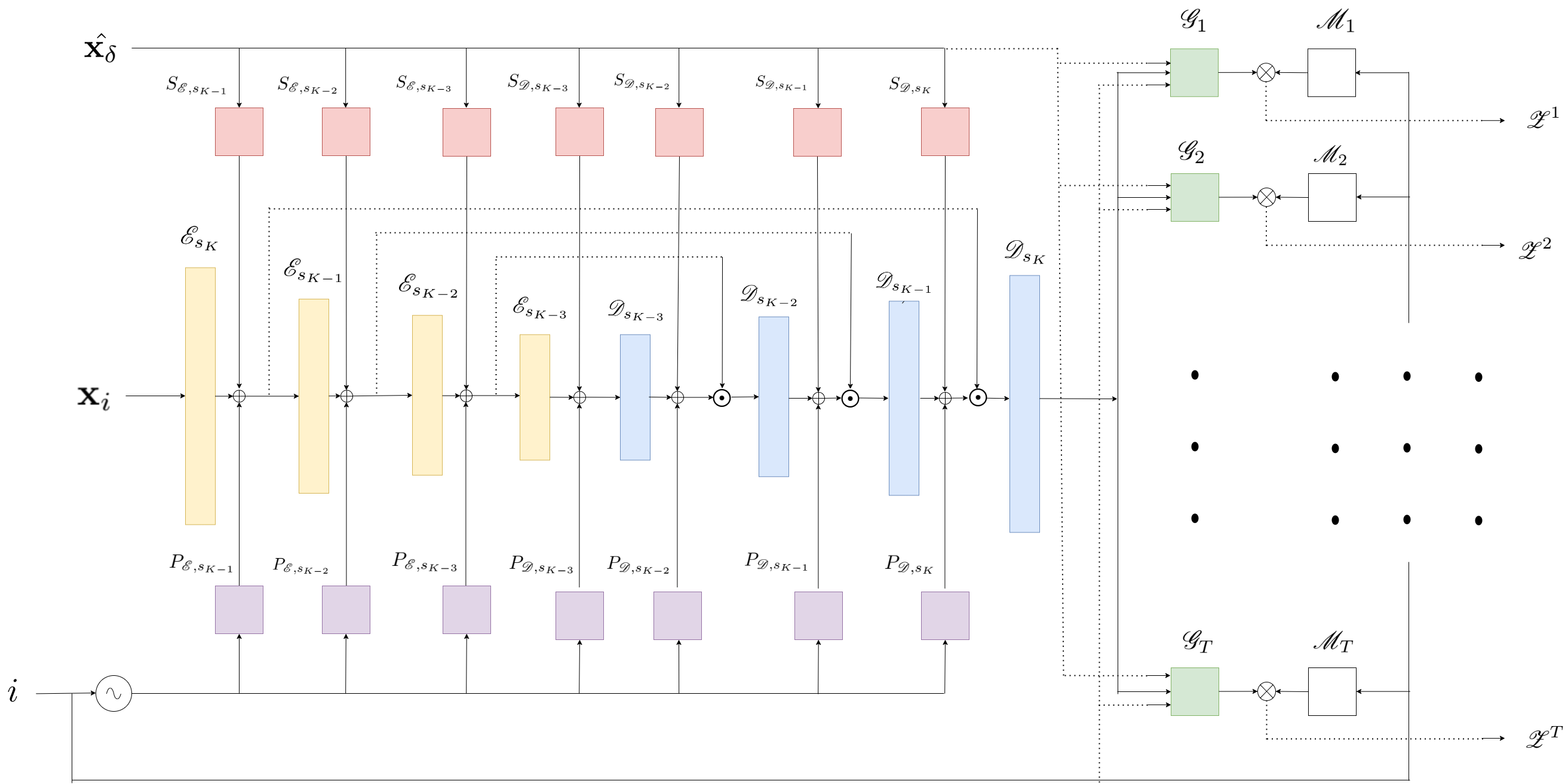}
\caption{Reverse diffusion model architecture}
\label{fig:model}
\end{figure}

The model, $r_{\phi}(\vb{x}_i,\hat{\vb{x}_{\delta}},i)$ takes the noisy image $\vb{x}_i$, its predicted scale $\hat{\vb{x}_{\delta}}$ and time step information $i$ as input and predicts additive noise for all the steps of the forward diffusion. Here $\phi$ is the set of trainable parameters of $r_{\phi}(\vb{x}_i,\hat{\vb{x}_{\delta}},i)$. These predictions can then be used to generate the reverse trajectory using Eqns. \ref{equ:posterior} and \ref{equ:reverse_samples_eps} with $\Tilde{\vb{\epsilon}_i}$ replaced by predictions of the model.

An encoder layer output, $Z_{\mathscr{E},s_k}$  is defined as:
\begin{equation}
    Z_{\mathscr{E},s_k} = \mathscr{E}_{s_{k+1}}(Z_{\mathscr{E},s_{k+1}},S_{\mathscr{E},s_{k+1}}(\hat{\vb{x}_{\delta}}),P_{\mathscr{E},s_{k+1}}(i);\phi)
\end{equation}
where the subscripts $s_{k}$ and $s_{k+1}$ represent respectively, the  output and input resolution levels. $\mathscr{E}_{s_{k+1}}(.)$ is the usual encoder layer in the U-Net comprising of Self-Attention \cite{ATT} Residual Blocks. $S_{\mathscr{E},s_{k}}(\hat{\vb{x}_{\delta}})$ is thus a feature mapping of $\hat{\vb{x}_{\delta}}$. $P_{\mathscr{E},s_{k+1}}(i)$ is a non-linear projection of the positional encoding of a time-step $i$.

Similarly, for all but the last layer of the decoder, the output, $Z_{\mathscr{D},s_{k+1}}$ is:
\begin{equation}
    Z_{\mathscr{D},s_{k+1}} = \mathscr{D}_{s_{k}}(Z_{\mathscr{D},s_{k}},Z_{\mathscr{E},s_k},S_{\mathscr{D},s_{k}}(\hat{\vb{x}_{\delta}}),P_{\mathscr{D},s_{k}}(i);\phi)
\end{equation}
where $\mathscr{D}_{s_{k}}$ is the usual decoder layer in the U-Net. 

The last layer of the decoder has $T$ feature maps. The $j^{th}$ feature map, $\mathscr{Z}^j$ is:
\begin{equation}
    \mathscr{Z}^j = \mathscr{M}_j\odot\mathscr{G}_j(Z_{\mathscr{D},s_{K}},\mathscr{P}(j),\mathscr{S}(\hat{\vb{x}}_{\delta});\phi_j),\,\,\, j \,\in \{1,\cdots,T\}
\end{equation}

where $\phi_j \subset \phi$. $\mathscr{M}_j$ is a channel mask which is all 1's if $j < i$, otherwise it is all 0's. This ensures that only predictions for time-steps preceding $i$ are made. $\mathscr{G}_j(.)$ is a feature map implemented using a small neural network. $\mathscr{P}(j)$ is a non-linear mapping of time-step $j$ with same dimensions as a single channel of $Z_{\mathscr{D},s_{K}}$ and $\mathscr{S}(\hat{\vb{x}}_{\delta})$ is a non-linear mapping of $\hat{\vb{x}}_{\delta}$. This mapping is fused with $Z_{\mathscr{D},s_{K}}$ before being fed to $\mathscr{G}_j(.)$. Exact details of the encoder and decoder layer structures are given in Appendix \ref{appendix:arch}.

The $T$ loss functions to minimize are similar to the one used by \cite{DDPM}:
\begin{equation}
    L(\phi,j) = \E_{i,x_i}[||\mathscr{Z}^j - \Tilde{\vb{\epsilon}_j}||_2^2],\,\,\,j \,\in \{1,\cdots,T\}
\end{equation}
For $j > i$, $L(\phi,j)$ is fixed to 0 as a consequence of the same argument of using the mask $\mathscr{M}_j$. $L(\phi,j)$ for different values of $j$ are optimized parallely. This way the parameters of the common network backbone:$\phi_b = \{\phi_l|\phi_l \in \phi, \phi_l \notin \phi_j,  \forall j\}$  are trained by all the $L(\phi,j)$, while the parameters $\phi_k$ of a particular 
 $\mathscr{G}_k(.)$  are trained only by its particular loss function $L(\phi,k)$.

The versatility of this model architecture is that it can be used for both denoising a noisy image as well as for generative tasks, depending on the type of autoencoder used for generating $\Tilde{\vb{x}}_{\delta}$.

Algorithm \ref{alg:sampling} in Appendix \ref{appendix:samp_alg} shows the sampling procedure. The algorithm is similar to the one used by \cite{DDPM}, the difference being that the scheduling parameters are calculated from $\hat{\vb{x}}_{\delta}$ and only one model run is required.

\section{Experiments and Results}
The models were trained on Cifar10 and CelebA datasets. The images were first normalized to the range (0,1] and a scalar value of $10^{-3}$ was added to all the pixels to ensure their values are greater than $0$. For denoising task, a denoising autoencoder \cite{DENAE} was trained, whereas for generative task, a VAE\cite{VAES} was used to estimate  $\vb{x}_{\delta}$. The reverse diffusion model for both the tasks is same. All the models are based on U-net architecture\cite{UNET}. Time-steps were encoded using sinusoidal positional embedding \cite{ATT}. The exact architecture details are provided in Appendix \ref{appendix:arch}.

\begin{figure}[t]
\centering
\begin{subfigure}[t]{0.49\textwidth}
    \centering
    \includegraphics[width=0.8\linewidth, height=4cm]{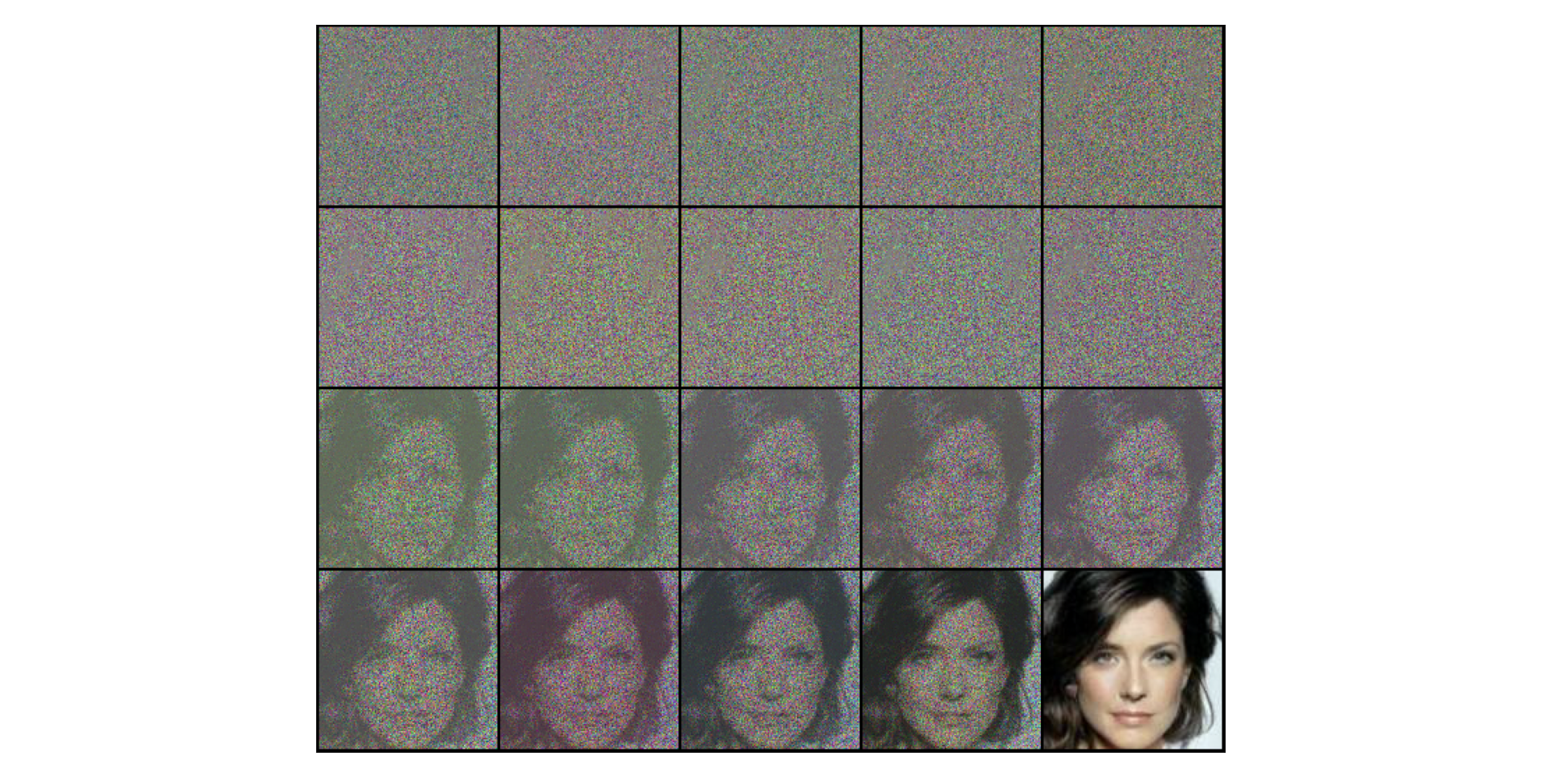}
    \caption{Generated image reverse trajectory}
    \label{fig:rev_traj_gen_model}
\end{subfigure}
\begin{subfigure}[t]{0.49\textwidth}
    \centering
    \includegraphics[width=0.8\linewidth, height=4cm]{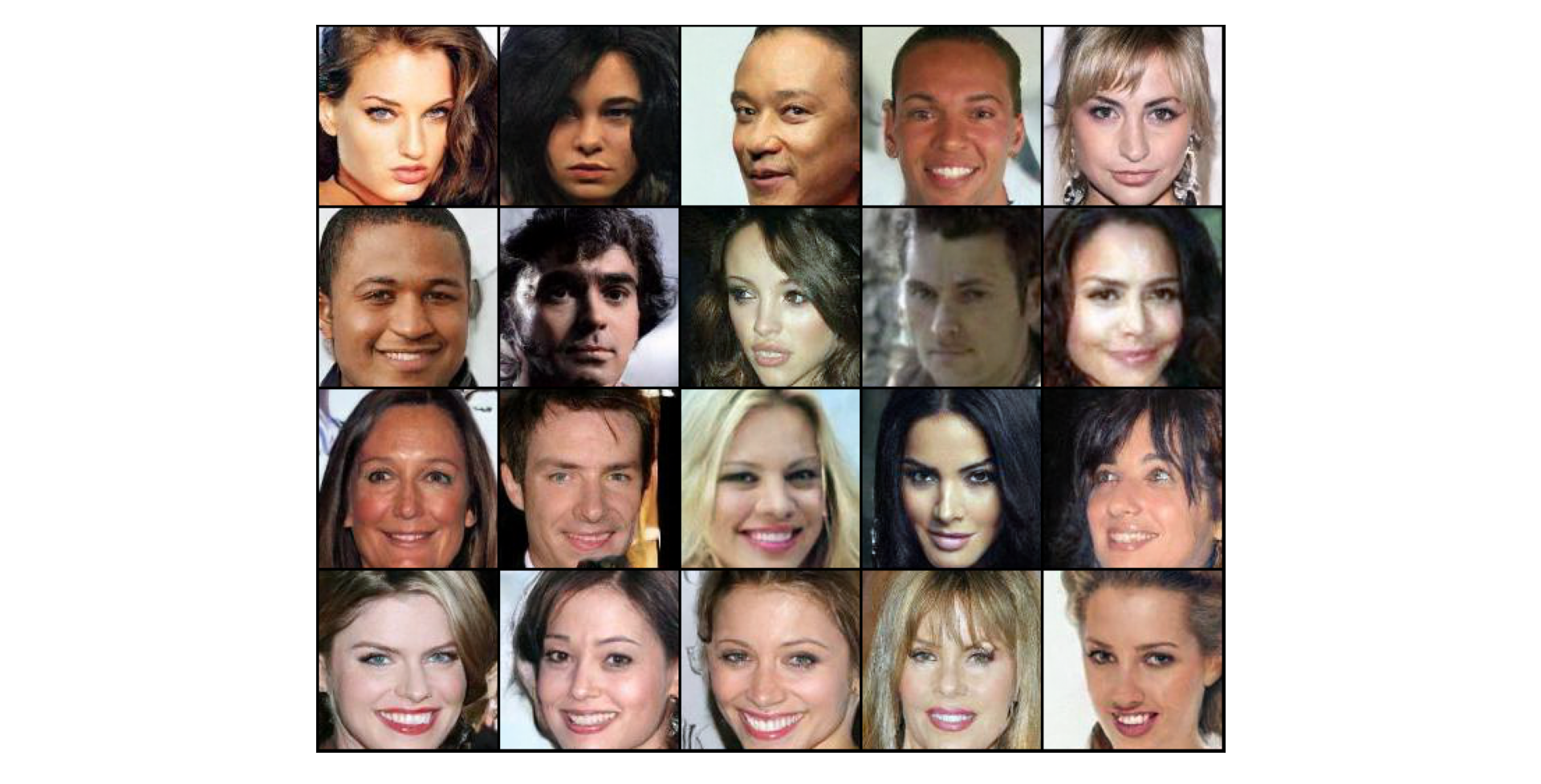}
    \caption{Image generation examples}
    \label{fig:rev_traj_gen_ex}
\end{subfigure}
\caption{Generated image reverse trajectory and image generation examples}
\end{figure}

Fig.  \ref{fig:rev_traj} shows the comparison of last 20 steps of reverse trajectories generated in case of image denoising. Fig.  \ref{fig:rev_traj_real} shows an example when we generate reverse trajectory by sampling from the probability distribution in Eqn. \ref{equ:posterior} by using real mean $\Tilde{\mu}(\vb{x}_i,\Tilde{\epsilon}_i)$ and real values of $\alpha_i$ and $\overline{\alpha}_i$. Fig.  \ref{fig:rev_traj_est} on the other shows reverse trajectory generated by relying on the posterior mean $\hat{\mu}(\vb{x}_i,\hat{x}_0)$, where $\hat{\vb{x}}_0$ is calculated directly from $\hat{\vb{x}}_{\delta}$ and so are $\alpha_i$ and $\overline{\alpha}_i$. We can see that estimation errors in the scheduling parameters and estimated clean image, $\hat{\vb{x}}_0$, result in poor image quality. However, when the reverse diffusion model is also used to refine the output by refining the mean of the posterior as $\Tilde{\mu}(\vb{x}_i, \mathscr{Z}^i)$, image quality is much better, as shown in Fig. \ref{fig:rev_traj_model}. Fig.  \ref{fig:rev_traj_gen_model} similarly shows example output trajectory from a generative model. Fig. \ref{fig:rev_traj_gen_ex} shows some generated examples. More examples can be found in Appendix \ref{Appendix:ex}

We compared our algorithm to DDPM and a SDE based model. Tables \ref{tab:cifargen} and \ref{tab:celebgen} show image generation performance of the three models on CIFAR10 and CelebA dataset respectively in terms of SSIM and execution time too. 

While image quality of our model is somewhere middling among the three models, the execution time (for denoising or image generation) is at least 4 times lesser than that of DDPM (our 0.3 second against their 1.23 seconds). DDPM requires at least 1000 steps to generate good quality images, even then it's performance is not as good as ours. The SDE based model effectively requires orders of magnitude more steps as a result of correction required due to MCMC subsampling. Our model on the other hand requires only 200 time-steps and 500 time-steps long trajectory to provide comparable performance, and even these steps of the trajectory are generated in a single run of the model.

\begin{figure}[t]
\centering
\begin{subfigure}[t]{0.3\textwidth}
\includegraphics[width=1\linewidth, height=2.5cm]{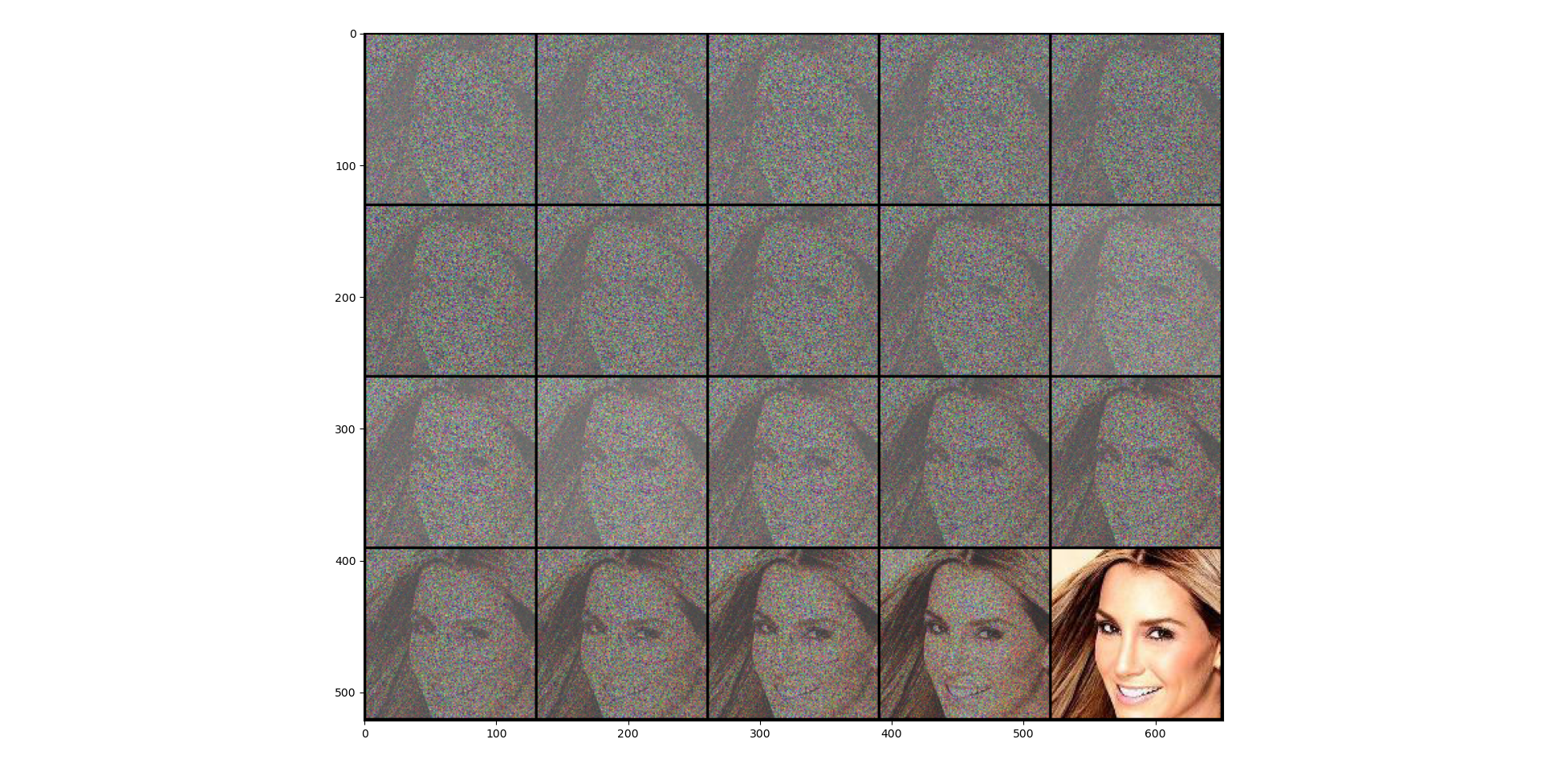}
\caption{Real trajectory}
\label{fig:rev_traj_real}
\end{subfigure}
\begin{subfigure}[t]{0.3\textwidth}
\centering
\includegraphics[width=0.69\linewidth, height=2.5cm]{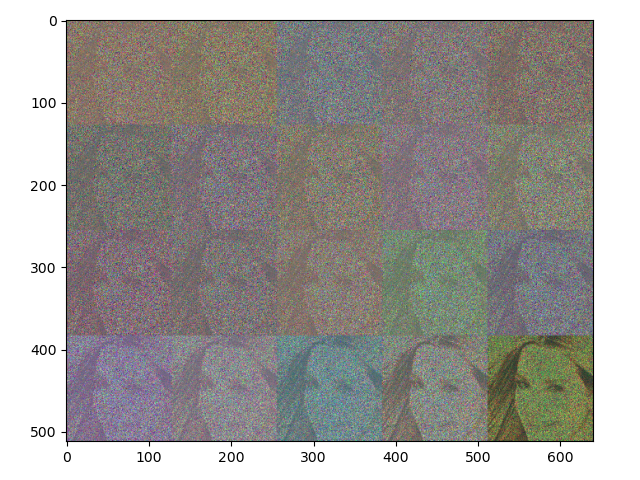}
\caption{Scale autoencoder}
\label{fig:rev_traj_est}
\end{subfigure}
\begin{subfigure}[t]{0.3\textwidth}
\centering
\includegraphics[width=1\linewidth, height=2.5cm]{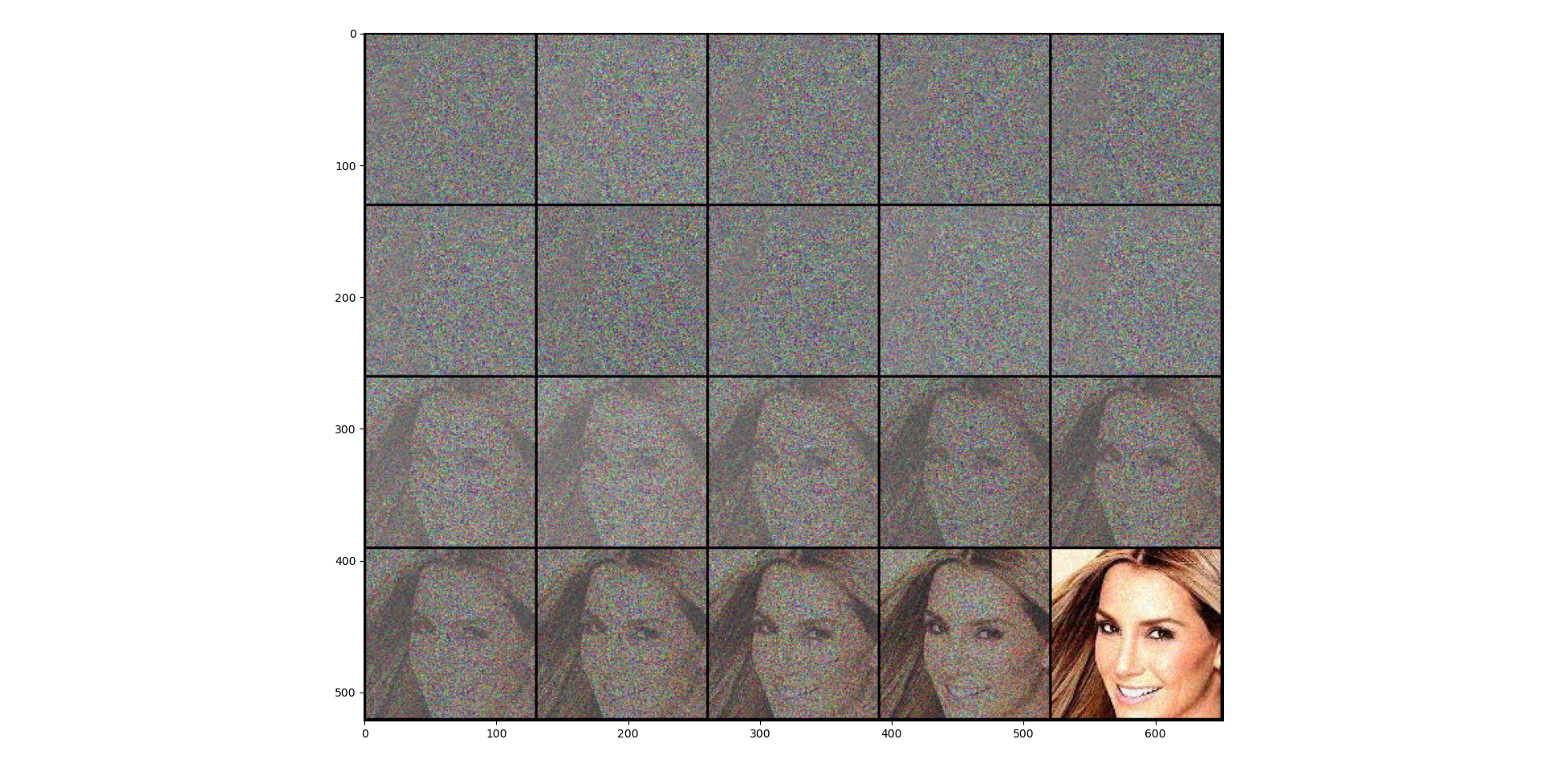}
\caption{Refined reverse diffusion model}
\label{fig:rev_traj_model}
\end{subfigure}
\caption{Reverse-time diffusion trajectory for i ranging from 20 to 1}
\label{fig:rev_traj}
\end{figure}

\begin{table}[!ht]
\begin{minipage}{.5\textwidth}
\begin{tabular}{@{}ccc@{}}
\hline
\multicolumn{3}{c}{CIFAR10 generative performance}                    \\ 
\hline
Model & FID & Execution Time (in seconds)   \\
DDPM     & 3.28      & 1.26    \\
SDE based     & 2.99     &   47.67  \\
Our Model & 3.15     & \textbf{0.3}    \\

\hline
\end{tabular}
\caption{}
\label{tab:cifargen}
\end{minipage}%
\hfill
\begin{minipage}{0.5\textwidth}
\begin{tabular}{@{}ccc@{}}
\hline
\multicolumn{3}{c}{CelebA generative performance}                    \\ 
\hline
Model & FID & Execution Time (in seconds)   \\
DDPM     & 3.51      & 10.19    \\
SDE based    & 3.20      &  246.69   \\
Our Model & 3.25      & \textbf{1.3 }   \\
\hline
\end{tabular}
\caption{}
\label{tab:celebgen}
\end{minipage}%
\end{table}

\section {Limitations and future work}
While our approach has an overall accuracy and time advantage over conventional models, a drawback is the hyperparameter nature of $\gamma$. Reducing the number of time steps $T$ of the diffusion process sets an upper bound on the choice of $\gamma$. Using a value greater than $T$ results in an exceedingly fast decay in the forward direction which breaks the Markovian condition, while using a smaller value results in non-convergence of some lower valued-pixels to standard normal distribution. Future work includes a plan for a systematic choice of the hyperparameter by jointly considering the data/pixels distribution for optimizing some associated energy functional while ensuring the existence of a stationary distribution as is commonly the case for a Markov chain. Additional experiments with much more diverse datasets are also planned for a more profound understanding of a universal selection of parameters.

A particularly interesting angle about this stochastic diffusion is to investigate the performance of our proposed diffusion by exploring the internal mechanics of the encoder and decoder in the context of diffusion processes. Our model learns the reverse diffusion process conditional on the global image structure (via $\vb{x}_{\delta})$. A future challenge is to discover the extent of control over a diffusion process for a system of interacting particles \cite{Bao:04, CONVDIFF} instead of parallel independent forward diffusions as is the norm in current models. Given the asymmetry of information processing between the encoder and decoder implementation of the U-Neural Network model uncovered in \cite{UNETROLE}, it would be interesting to study what kind of semantic information is learned by the two subcomponents of the model. Finally, we also intend to study how our approach works on transformer based models to improve performance.

\section{Conclusion}
In this paper, we have introduced a novel forward diffusion model that significantly improves upon the limitations of conventional models in terms of convergence speed and computational efficiency. By leveraging the microscopic structure of clean images to learn the drift and diffusion coefficients, our model degrades the Signal to Noise Ratio much faster than traditional approaches. Drawing inspiration from the water pouring algorithm, we implemented a pixel-based scheduling strategy that optimizes forward diffusion by considering the initial values of individual pixels. This method achieves isotropic Gaussian distribution across the pixels more efficiently than conventional, pixel-agnostic diffusion methods.

Furthermore, we utilized an autoencoder to learn a comprehensive diffusion schedule. The learned knowledge of the global structure of the clean image inspired us to develop a reverse-time data driven diffusion model to generate the entire reverse-time diffusion trajectory in one step. This approach not only maintained image quality but also accelerated the reverse-time diffusion process by up to 10 times compared to existing models. Our findings demonstrate that addressing the inefficiencies of universal diffusion in generative models through a detailed, pixel-focused approach can lead to substantial improvements in performance, paving the way for more effective and efficient generative image modeling.

\newpage
\bibliographystyle{plainnat}
\bibliography{references}

\newpage
\appendix

\section{Proof that exponential pixel value based diffusion follows water pouring algorithm}
\label{Appendix:snr}


In this section, we will show that in each step higher variance noise is being added to a higher value pixel in comparison to a lower value pixel.

SNR of pixel $j$ at time $t$ is calculated as 
\begin{equation}
    \text{SNR}(j,t) = \dfrac{(x^j_0)^2}{e^{ \gamma\,\,t\,\, x^j_0}-1}
\end{equation}

Thus, the rate of change of SNR over time $t$ is
\begin{equation}
    \dfrac{d\,\text{SNR}(j,t)}{dt} = -\dfrac{\gamma(x^j_0)^3}{(e^{ \gamma\,\,t\,\, x^j_0}-1)^2}
        \label{equ:appendixa0}
\end{equation}
From the Taylor series
\[ e^x = 1+ x + \frac{x^2}{2!} + \cdots \]
 we can show that for any small value $x >0$, the following inequality is true. 
\[ 0< xe^{x/2} < e^{x} -1 < xe^x \]
This leads to 
\[ 0<\frac{1}{x e^{x}} < \frac{1}{(e^{ x} -1)}  <\frac{1}{x e^{x/2}}  \]

Applying above inequality to our case, we can see that 
SNR at time $t$ can be expressed as 
\begin{equation}
     0<\frac{x^j_0}{( \gamma\,\,t\,\,) e^{ \gamma\,\,t\,\, x^j_0}} < \dfrac{(x^j_0)^2}{e^{ \gamma\,\,t\,\, x^j_0}-1}  <\frac{x^j_0}{( \gamma\,\,t\,\,)e^{ \gamma\,\,t\,\, x^j_0/2}}   
    \label{equ:appendixa1}
\end{equation}   
And the magnitude of rate of change of SNR with time, $t$ can be expressed as 
\begin{equation}
     0<\frac{x^j_0}{\gamma\,\,t^2\,\,  e^{2 \gamma\,\,t\,\, x^j_0}} < \dfrac{\gamma(x^j_0)^3}{(e^{ \gamma\,\,t\,\, x^j_0}-1)^2}  <\frac{x^j_0}{\gamma\,\,t^2\,\,e^{ \gamma\,\,t\,\, x^j_0}}  
    \label{equ:appendixa2}
\end{equation}

From Eqns.  \ref{equ:appendixa1} and \ref{equ:appendixa2} we can see that for $t$ belonging to a small interval, $t \in [0, t_{\delta}]$, both the SNR as well as the magnitude of its rate of change of SNR at pixel $x^j_0$ are very high and it quickly approaches $0$. In this small interval $ [0, t_{\delta}]$, the SNR decays  at the rate of $t^{-2}$. The decay rate is proportional to $x^j_0$ also. Therefore, when $0<x^j_0 < x^k_0$, in the interval 
$t \in [0, t_{\delta}] $, we have

\section{Proof of faster convergence of our forward diffusion process}
\label{Appendix:rates}
For any discrete diffusion process satisfying, Eqn. \ref{equ:iterative_fwd}, the corresponding continuous SDE is 
\begin{equation}
    d\vb{x}_{t} = -\dfrac{\beta_t}{2}\vb{x}_tdt + \sqrt{\beta_t}d\textbf{w},\,\,\, t \in (0,1)
\end{equation}

Thus, the drift and diffusion terms in Eqn. \ref{equ:continous} can be expressed as $\vb{f}(\vb{x}_t,t) =  -\dfrac{\beta_t}{2}\vb{x}_t$ and $g(\vb{x_t},t) = \sqrt{\beta_t}$. From theory of Ito Calculus \cite{SARKA_SDE, Ito, Geometrydiffusion}, 

\[  x_t- x_0- \int_0^t \vb{f}(\vb{x}_s,s) ds=  x_t- x_0-\int_0^t -\dfrac{\beta_s}{2}\vb{x}_s ds  \]
is a martingale (for definition see \cite{Ito}). Thus  we can calculate rates of change of mean of the image at $t$ as:
\begin{equation}
    \dfrac{d\,\E[x_t]}{dt} = \E[\vb{f}(\vb{x}_t,t)] = -\dfrac{\beta_t}{2}\E[x_t] 
    \label{equ:DDPM_mean_rate}
\end{equation}
This leads to 
\[
\E[x_t] = \mathlarger{x_0} \mathlarger{e}^{-\bigintss_0^t \dfrac{\beta_s}{2}\,ds}
\]
To focus on analyzing the algorithm, we simply let $\beta_t =at$ (a linear function of time). Hence the expected trajectory of the DDPM follows the exponential decay of the form
\begin{equation}
    E[x_t] = x_0e^{-\frac{at}{2}^2}.
    \label{equ:DDPM_exp_traj}
\end{equation}    
We now compare the expected trajectory of our model with the trajectory in Eqn \ref{equ:DDPM_exp_traj}

The continuous SDE corresponding to our pixel-wise diffusion given in \ref{Proposition2} is:

\begin{equation}
dx^j_t =-\frac{\gamma x^j_0}{2}x^j_t dt + \sqrt{\gamma\,x^j_0} dw^j,
\end{equation}
This SDE can be derived from the discrete iterative forward diffusion equation of our process using the same approach as described in \cite{Scored}.

In our case, the rate of change of mean turns out to be:
\begin{equation}
    \dfrac{d\,\E[x_t^j]}{dt} = -\frac{\gamma x^j_0 }{2}\E[x_t^j] 
    \label{equ:exp_mean_rate}
\end{equation}



This leads to 
\begin{equation}
    \E[\vb{x}_t] = \vb{x}_0 \odot e^{- \frac{\gamma\vb{x}_0 t}{2}}
    \label{equ:our_exp_traj}
\end{equation}

By further choosing $\gamma$ in such a way that $\gamma x^j_0 > at ,\,\,\,\forall\,\,j\in \{1,\cdots,d\},\,\,\, t \in (0,1]$, we can ensure that the trajectory in Eqn \ref{equ:our_exp_traj} decays faster than the DDPM trajectory in Eqn \ref{equ:DDPM_exp_traj}.

This demonstrates that our diffusion accounts for the pixel value in deciding the rate of decay in the forward process, which in turn requires a selection of a specific value of $\gamma$ as described in Section \ref{sec:def} to ensure effective pixel-wise diffusion. 

The drift term of each pixel diffusion can be further generalized to $-\Gamma( x^j_0, t)x^j_0$,  where $\Gamma( x^j_0, t)$  is a positive function of $t$ and $x^j_0$, monotonic in $t$. Similarly, Over all the pixels, we can have a vector valued function $\boldsymbol{\Gamma}( \vb{x}_0, t)  \in \R^d$.

For comparison with conventional DDPM, one can select $\boldsymbol{\Gamma}( \vb{x}_0, t) =\gamma\frac{\beta_t}{2} \vb{x}_0$, where $\beta_t$ is the same parameter used in conventional DDPM. This makes the diffusion schedule pixel-dependent. If $\beta_t = at$, we will have
\begin{equation}
    \E[\vb{x}_t] = \vb{x}_0 \odot {e}^{- \frac{\gamma a \vb{x}_0 t^2}{2}}
\end{equation}

This demonstrates the efficiency of our pixel-wise modulated diffusion has over  the DDPM algorithm.

\section{Model architecture and other details}
\label{appendix:arch}
Both the autoencoder as well as the reverse diffusion model, were based on U-Net architecture. However, as autoencoder training involved learning simpler image structures, the depth of the network was kept much lower. Each autoencoder encoder/decoder composed of a single residual convolutional block along with self attention of $16 \times 16$ context size followed by a downsampling/upsampling by an order of 2. Group normalization was applied to the input of each convolutional layer. The time steps were encoded using Sinusoidal positional embedding with encoding vector size chosen as 128. In case of denoising autoencoder, the bottleneck layer was composed of feature maps of $4 \times 4$ resolution. In case of variational autoencoder the $4 \times 4$ flattened and projected to a larger size latent space to ensure generation diversity.

In case of reverse diffusion model, each encoder/decoder layer consisted of 2 residual convolutional blocks with self attention. The same time encoding employed by the autoencoders was also used by the reverse diffusion models. While the number of channels in the intermediate layers differed according to the underlying dataset, the number of channels in the penultimate layer were kept fixed at 512. The final layer involved parallel sub-networks to model various time step outputs. Each sub-network was modelled as 2 successive residual blocks with self attention with number of output channels being same as the input image (i.e. 3 in case of RGB images). Adam optimizer was used for backpropagating through all the networks. The learning rates used for CIFAR10 and CelebA database were respectively $10^{-4}$ an $2 \times 10^{-4}$ respectively, with training being done over 1.5 M iterations. The batch sizes for the two datasets were respectively 128 and 16. Finally the number of time steps of the diffusion process were fixed to 200 and 500 respectively. Training and inference was done on a single NVIDIA Tesla V100 SXM2 32 GB GPU. The same GPU was used to compare execution time of conventional models.

\section{Algorithm for sampling from the trained reverse diffusion model}
\label{appendix:samp_alg}
\begin{algorithm}[!h]
    \caption{Sampling algorithm}
  \begin{algorithmic}[1]
    \REQUIRE Pretrained scale autoencoder $s_{\theta}(.)$ and reverse diffusion model, $r_{\phi}(.)$.
    \INPUT Noisy image $\vb{x}_i$ and time-step $i$ of the forward diffusion
    \OUTPUT Clean image $\hat{\vb{x}_0}$
    \STATE Scale estimate: $\hat{\vb{x}_{\delta}}$ = $s_{\theta}(\vb{x}_i,i)$
    \STATE ${\boldsymbol{\alpha}}_j = \exp{(\dfrac{1}{T}\log{(\hat{\vb{x}_{\delta}})})},\,\,\,\forall j \in {1,\cdots,T}$
    \STATE $\overline{\boldsymbol{\alpha}}_j = \exp{(\dfrac{j}{T}\log{(\hat{\vb{x}_{\delta}})})},\,\,\,\forall j \in {1,\cdots,T}$
    \STATE $\Tilde{\boldsymbol{\beta}}_j = \dfrac{\bf{1}-\overline{\boldsymbol{\alpha}}_{j-1}}{\bf{1}-\overline{\boldsymbol{\alpha}}_j}(\bf{1}-\boldsymbol{\alpha_j}),\,\,\,\forall j \in {1,\cdots,T}$
    \STATE Reverse diffusion noise predictions: $\mathscr{Z} = r_{\phi}(\vb{x}_i,\hat{\vb{x}_{\delta}},i)$
    \STATE \textbf{Initialization} $j = i,\,\,\,\hat{\vb{x}}_j = \vb{x}_i$ 
    \WHILE{$j > 0$}
        \STATE $\vb{z} \sim \mathcal{N}(\textbf{0},\textbf{I})$
        \STATE $\hat{\vb{x}}_{j-1} = \dfrac{\bf{1}}{\sqrt{\boldsymbol{\alpha}_j}}\odot\Bigg(\hat{\vb{x}}_j - \dfrac{\bf{1}-\boldsymbol{\alpha_j}}{\sqrt{\bf{1}-\overline{\boldsymbol{\alpha}}_j}}\odot\mathscr{Z}^j\Bigg) + \sqrt{\Tilde{\boldsymbol{\beta}}_j}\odot\vb{z}$
        \STATE $j = j - 1$
    \ENDWHILE
    \STATE \textbf{return} $\hat{\vb{x}}_{0}$
    \end{algorithmic}
    
    \label{alg:sampling}
\end{algorithm}

\section{Real and learned scheduling parameters comparisons}
\label{Appendix:scale_fig}
\begin{figure}[ht]
\centering
\begin{subfigure}[t]{0.48\textwidth}
\includegraphics[width=0.5\linewidth, height=4cm]{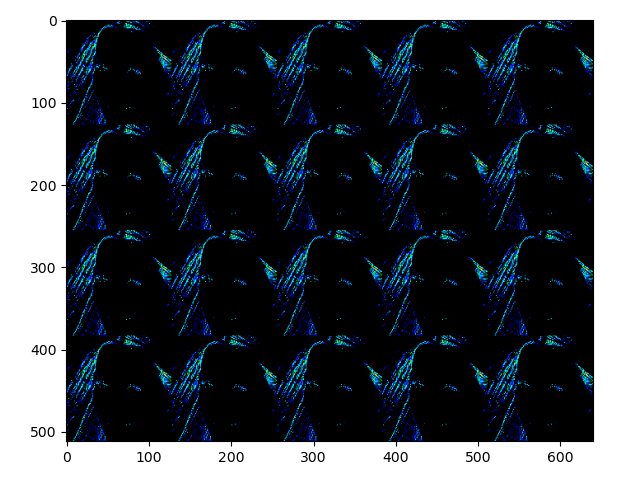}
\caption{Real image scale, $\vb{x}_{\delta}$.}
\label{fig:scale_real}
\end{subfigure}
\begin{subfigure}[t]{0.48\textwidth}
\centering
\includegraphics[width=0.5\linewidth, height=4cm]{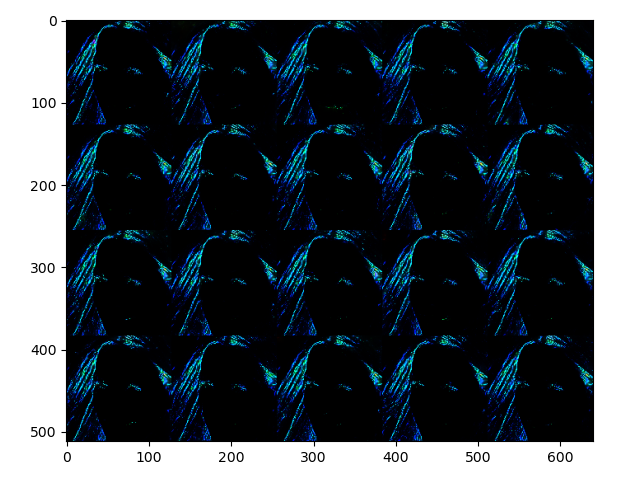}
\caption{Image scale estimate, $\hat{\vb{x}}_{\delta}$.}
\label{fig:scale_est}
\end{subfigure}
\caption{Real and estimated image scales (estimated using the autoencoder using the noisy image $\vb{x}_i$) for i ranging from 480 to 500.  Note that the real image scale $\vb{x}_{\delta}$ is same for all $i$.}
\label{fig:scale_comp}
\end{figure}

\newpage
\begin{figure}[ht]
\centering
\includegraphics[width=1\linewidth, height=7cm]{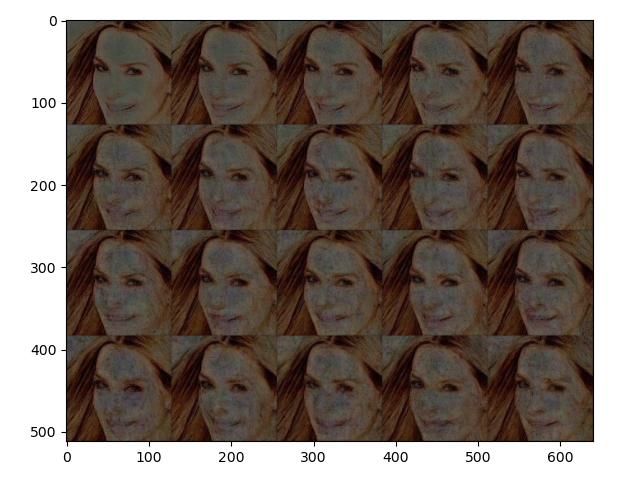}
\caption{Estimate of the clean image $\vb{x}_0$ from the estimated image scale $\hat{\vb{x}}_0$ for i ranging from 1 to 20}
\label{fig:img_from_img_delta}
\end{figure}
\newpage

\section{Generated image examples}
\label{Appendix:ex}
\begin{figure}[ht]
\centering
\includegraphics[scale=0.7]{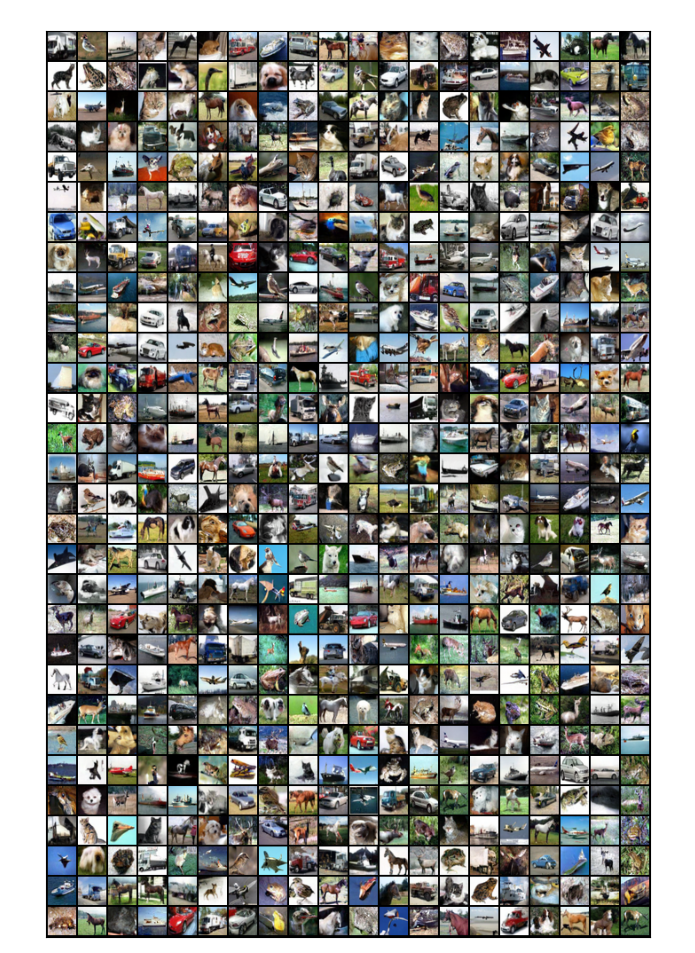}
\caption{Cifar10 examples}
\label{fig:img_cifar_ex}
\end{figure}

\begin{figure}[ht]
\centering
\includegraphics[scale=0.63]{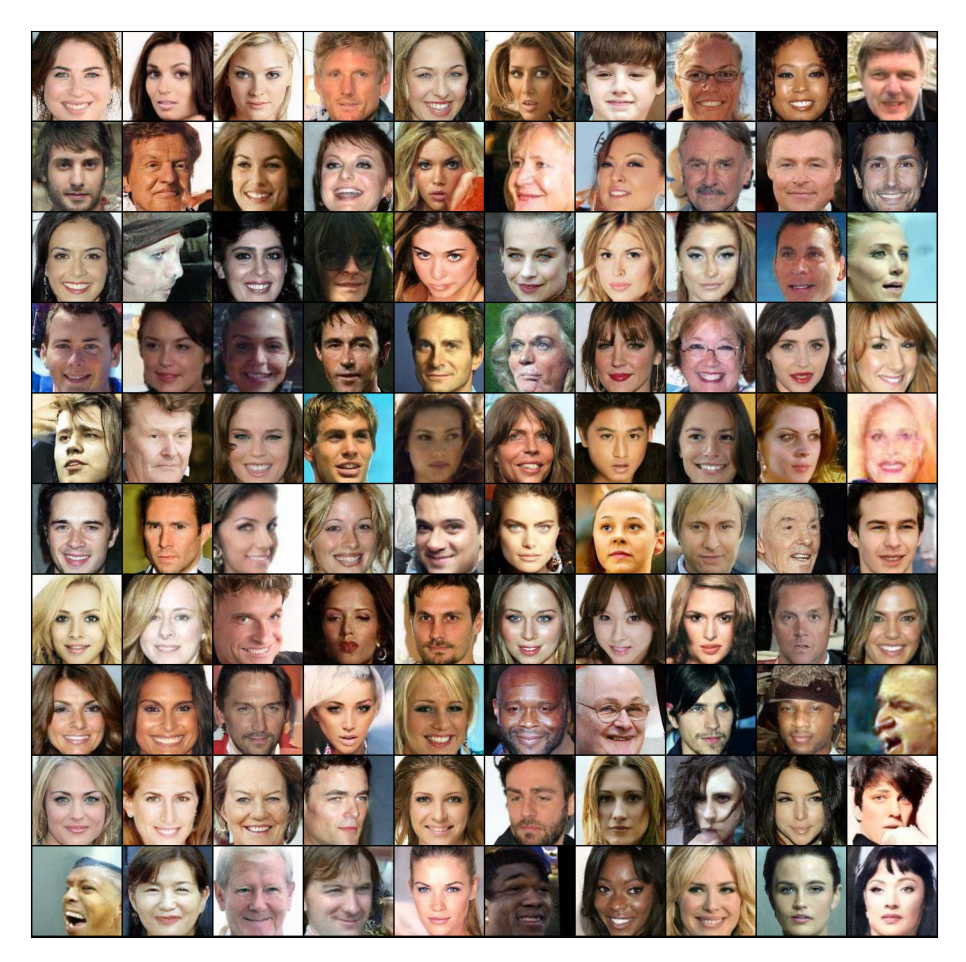}
\caption{CelebA examples}
\label{fig:img_celeba_ex}
\end{figure}

\end{document}